\documentclass[aps,prx,reprint,superscriptaddress,amsmath,amssymb,longbibliography,nofootinbib]{revtex4-2}

\usepackage{graphicx}
\usepackage{booktabs}
\usepackage{xcolor}
\usepackage[normalem]{ulem}
\usepackage{amsmath,amssymb,mathtools}
\usepackage{hyperref}
\hypersetup{colorlinks=true,linkcolor=blue!50!black,citecolor=blue!50!black,urlcolor=blue!60!black}

\newcommand{\lean}{\textsc{Lean}}
\newcommand{\Tr}{\operatorname{Tr}}
\newcommand{\epsqkd}{\varepsilon_{\mathrm{QKD}}}
\newcommand{\epsAT}{\varepsilon_{\mathrm{AT}}}
\newcommand{\epsEV}{\varepsilon_{\mathrm{EV}}}
\newcommand{\epsPA}{\varepsilon_{\mathrm{PA}}}

\newcommand{\lcbrk}{\discretionary{}{}{}}
\newcommand{\leanlink}[2][]{%
  \begingroup\renewcommand{\_}{\textunderscore\lcbrk}%
  \if\relax\detokenize{#1}\relax
    \texttt{#2}%
  \else
    \href{#1}{\texttt{#2}}%
  \fi\endgroup}
\newcommand{\leanchor}[2][]{\par\nobreak\vspace{1pt}{\raggedleft\footnotesize\sloppy
  \lean{}:\ \leanlink[#1]{#2}\par}%
  \noindent\ignorespaces}

\renewcommand{\thesubsection}{\thesection.\Alph{subsection}}
\makeatletter\renewcommand{\p@subsection}{}\makeatother

\begin{document}

\title{Quantum Cryptography and Autoformalization in Lean}

\author{Maor Ben-Shahar} \email{maorbs@mit.edu} \affiliation{MIT Center for Theoretical Physics -- a Leinweber Institute, Cambridge, Massachusetts 02139, USA} \author{Jonathan Morag} \affiliation{Research Laboratory of Electronics, Massachusetts Institute of Technology, Cambridge, Massachusetts 02139, USA} \author{Dirk Englund} \email{englund@mit.edu} \affiliation{Research Laboratory of Electronics, Massachusetts Institute of Technology, Cambridge, Massachusetts 02139, USA} \affiliation{Department of Electrical Engineering and Computer Science, Massachusetts Institute of Technology, Cambridge, Massachusetts 02139, USA} \author{Kfir Sulimany} \email{kfir@mit.edu} \affiliation{Research Laboratory of Electronics, Massachusetts Institute of Technology, Cambridge, Massachusetts 02139, USA}

\date{\today}

\begin{abstract}
Quantum key distribution provides secure communication by relying on physics principles and mathematical reasoning. Security proofs across the field can contain gaps, often between individually valid theorems integrated under mismatched assumptions. We produce an end-to-end security theorem formalized in the \lean{} proof assistant, eliminating hidden hypotheses by building only on \lean{}'s standard axioms. Our proof covers composable security of the entanglement-based qubit BB84 protocol against coherent attacks by the postselection technique, accounting for finite block size, and is written in a language that enforces local operations and classical communication to help ensure a faithful representation of the protocol. The proof itself was enabled by LLM-based autoformalization, and with the proof complete, we used the same pipeline as a discovery machine, reducing the finite-size penalty on the key rate. Our work sets a reproducible standard for security arguments in quantum cryptography and shows how formal proof engines can accelerate discovery across quantum information science.

\end{abstract}

\maketitle

\section{Introduction}
\label{sec:intro}

Quantum key distribution (QKD) aims to establish shared secret keys for private communication, with their security derived from quantum theory under explicit assumptions about the devices and with authenticated classical communication.
Advances in protocols~\cite{WangTFQKD19}, experimental platforms~\cite{WeiMDI20}, and networks~\cite{TangMDImetro16} have widened the range of proposed and demonstrated QKD systems.
A central advance was composable security, which formalizes when a generated key can safely replace an ideal secret key in subsequent cryptographic applications~\cite{RennerKonig05,PortmannRenner21}.

A complete security proof must use compatible assumptions throughout.
For example, Nahar \emph{et al.}~\cite{Nahar24} closed a gap first identified by Belzig~\cite{Belzig24} in the application of postselection~\cite{CKR09} to QKD, obtaining a weaker security bound than previously claimed.
A recent comprehensive review~\cite{TTNKL-review} documents gaps in security proofs for practical decoy-state implementations of the Bennett--Brassard 1984 (BB84) protocol~\cite{BB84}, including incompatible assumptions when combining results from different papers. The review also identifies recent end-to-end proofs~\cite{Tupkary26,Mizutani25}.

In \lean~4~\cite{Lean4}, a formal proof assistant, a small trusted kernel checks that each proof establishes its formal statement from the declared hypotheses and axioms. When a proof uses an earlier theorem, the kernel checks that its hypotheses are satisfied.
Our review checks that the definitions and theorem statement describe the intended physical protocol and security claim.

Machine-checked security proofs are established in classical cryptography~\cite{EasyCrypt,CryptHOL}.
In quantum information, formal methods have been used to verify protocols and circuits~\cite{BoenderCoq15,SQIR,PengShor23} and develop reasoning frameworks for quantum programs~\cite{CoqQ,BordgIsabelle21}.
For QKD, formal verifications of BB84 have used process calculi, Coq and model checking~\cite{FMQuantumSurvey}. Nagarajan \emph{et al.}~\cite{PRISMQKD} model-checked finite, noiseless instances of BB84 under intercept--resend attacks, Kubota \emph{et al.}~\cite{Kubota16} verified protocol-equivalence steps in the Shor--Preskill proof, and Ax-Prover~\cite{AxProver} formalized a lemma of the Lo--Chau argument. None of these works gives an end-to-end composable finite-key security bound against general attacks.

Large language models (LLMs) coupled to proof assistants are now used in research mathematics, including the \lean{} formalization of Fermat's Last Theorem~\cite{FermatFormalization26} and a proposed proof of finite-time blowup for smoothly forced three-dimensional Navier--Stokes flow~\cite{NavierStokes26}.
Libraries such as Mathlib supply reusable mathematical definitions and results~\cite{mathlib}; formalizing QKD also requires definitions of permitted local operations, public communication, and a security criterion.

We present, to our knowledge, the first end-to-end machine-verified composable finite-key security proof of entanglement-based qubit BB84 against coherent attacks (Fig.~\ref{fig:stack}).
We model Alice's and Bob's operations in a language, defined in \lean{}, that enforces local operations and classical communication (LOCC)~\cite{ChitambarLOCC14} by construction.
For this explicit protocol, the proof combines the BB84 phase-error analysis~\cite{ShorPreskill00,Renner08} with privacy amplification~\cite{RennerKonig05}, finite-size entropy bounds~\cite{Tomamichel16,DupuisFawzi19}, and the postselection reduction that extends security against collective attacks to coherent attacks~\cite{CKR09,Nahar24}.
The audit described in Sec.~\ref{sec:method} checks that the definitions used by this theorem match the intended local operations and the information available to Eve.
We also bound the protocol's abort probability under a memoryless honest channel.

We prove a tighter bound on the conditional-entropy variance of classical--quantum states than that of Ref.~\cite{DupuisFawzi19}, reducing the finite-size penalty in the key length.
The improved bound and the key-length condition built on it are proved in \lean{}.

We release the underlying \lean~4 quantum-information library as open source so that its definitions and results can be reused beyond this BB84 proof.
It provides states and channels based on the standard finite-dimensional formalism~\cite{Watrous18}, together with entropies, finite-size entropy bounds, and the LOCC protocol language.
The same infrastructure has also been used to prove a quantum-optimization conjecture in \lean{}~\cite{FGGmachine}.
Our library complements existing formalizations of quantum information and physics, including \textsc{Lean-QuantumInfo}, now part of \textsc{Physlib}~\cite{LeanQuantumInfo,PhysLean}, as well as other works on quantum error correction~\cite{LeanQECRepos}, quantum algorithms and circuits~\cite{LeanAlgRepos}, and the foundations of quantum mechanics~\cite{LeanFoundationsRepos}.

Section~\ref{sec:method} describes the AI pipeline and the audit of definitions and theorem statements.
Section~\ref{sec:lean-locc} introduces the LOCC language and the security, robustness, and classicality definitions.
Section~\ref{sec:bb84-realization} constructs entanglement-based qubit BB84 and proves its security and robustness bounds.
Section~\ref{sec:discovery} presents the sharpened entropy-variance bound used by that guarantee and its finite-key consequences.

\section{The AI proving pipeline and review}
\label{sec:method}
\label{sec:manual-audit}

\begin{figure*}[t]
\centering
\includegraphics[width=\textwidth]{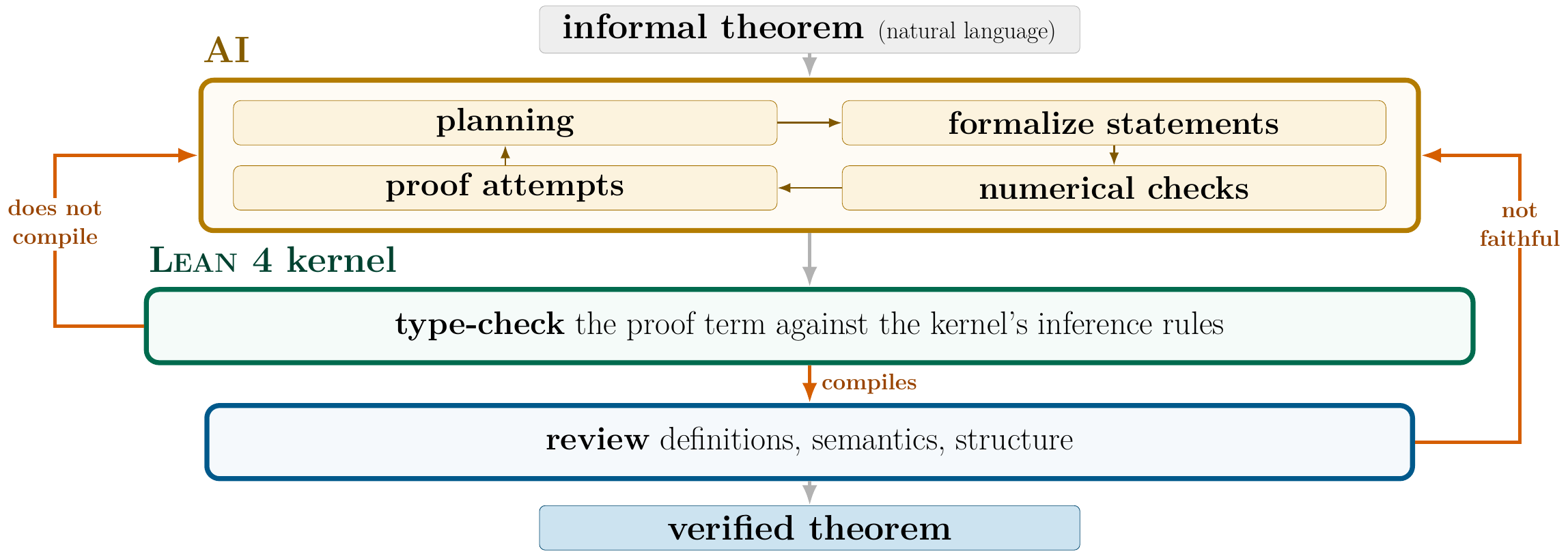}
\caption[The proving workflow and review.]{\textbf{The proving workflow and review.} Agents develop statements, definitions, and candidate proofs. Kernel checking validates the formal derivation, while manual review checks the intended meaning of the formal objects and guides proof strategy and library organization.}
\label{fig:pipeline}
\end{figure*}

We developed the formalization through an iterative collaboration between researchers and LLM agents, using \lean{} to check the resulting proofs (Fig.~\ref{fig:pipeline}). Agents based on Claude and GPT translated statements from target papers into \lean{}, guided by type-checking feedback from the \lean{} compiler. They decomposed proofs into intermediate claims and developed the necessary definitions and supporting results based on \textsc{Mathlib}~\cite{mathlib}. The agents reviewed candidate statements and used numerical checks, where applicable, to search for counterexamples or test proposed bounds before attempting formal proofs.
For sustained work we used our \href{https://github.com/maorbenshahar/loopy-prover}{\texttt{loopy}} sequencer to launch agents with task-specific context, including dependency graphs for the library from \href{https://github.com/maorbenshahar/lean-graph-tools}{\texttt{lean-graph-tools}}, both of which are released open source. We also ran a locally hosted GLM-5.3-Flash model~\cite{GLM53Flash} through \texttt{loopy} on proving tasks set by Claude and reviewed by Claude and GPT (Sec.~\ref{sec:disc}).

We guided the development of definitions and types and redirected overly specialized proof attempts toward general arguments. For example, we encouraged a proof of the postselection reduction for quantum channels~\cite{CKR09} in place of one specialized to the BB84 representation then in use. We also reviewed the organization and naming of definitions and results to make the library easier to navigate and reuse, and checked agreement with the cited literature.

The \lean{} kernel type-checks each proof term against its theorem statement~\cite{Lean4}. Our review can therefore focus on whether the statement, its hypotheses and its underlying definitions express the intended claim, without rechecking the proof body.
This review has caught formalization errors. For example, an earlier AI-generated protocol computed the key as a hash of Alice's and Bob's key strings combined, although neither party had access to both. This error motivated our development of the LOCC language introduced in Sec.~\ref{sec:lean-locc}, which enforces locality and makes public communication explicit.

\section{The LOCC language and security definitions}
\label{sec:lean-locc}

This section develops the security definition for a protocol-independent representation of QKD as an LOCC program, in which
steps are constructed as Kraus maps acting separately on each party's register. Composing these maps along each branch and summing over branches gives a channel whose output records every announcement in a public register.
The BB84 protocol is constructed and proved secure in Sec.~\ref{sec:bb84-realization}.
Throughout this and subsequent sections we link to our \lean{} library documentation below the relevant text.
\leanchor[https://maorbenshahar.github.io/QCryptLean/docs/]{QCryptLean}
Appendix~\ref{sec:glossary} provides a glossary of the main formal definitions used in the paper.

All our Hilbert spaces are finite-dimensional. Entropies are measured in bits,
$\log_2$ denotes the base-two logarithm and $\ln$ the natural logarithm.
We follow the notation of \lean{}, in which every object has a type. We write $x:T$, read ``$x$ has type $T$''. In set notation one would write $x\in T$. Types can depend on objects, for example $\mathsf{Prog}(\Gamma)$, defined below, is the type of programs starting from the system $\Gamma$, so $q:\mathsf{Prog}(\Gamma)$ says that $q$ is such a program.

\subsection{LOCC Program}

We enforce LOCC constructively by defining a type called Program whose inhabitants can only be constructed by composing local actions, potentially dependent on public classical communications. Before introducing this
Program, we specify the multipartite system and the private and announced actions from which it is constructed. Each action is built from an instrument on one party's local register, following the instrument description of LOCC in Ref.~\cite[Sec.~II]{ChitambarLOCC14}.

We consider a finite set of parties $P$ (Alice and Bob for QKD), each with a finite register basis label set $\Gamma_i$. We define the family $\Gamma$ that specifies the channel's multipartite input system:
\begin{equation}
\Gamma=(\Gamma_i)_{i:P},\qquad
\mathcal H_\Gamma=\bigotimes_{i:P}\mathbb C^{\Gamma_i}.
\end{equation}
\leanchor[https://maorbenshahar.github.io/QCryptLean/docs/QCryptLean/LOCC/MultipartiteSystem.html\#LOCC.MultipartiteSystem]{MultipartiteSystem}
An instrument $I$ for party $i$, with finite-dimensional input and output
spaces $\mathcal H_i$ and $\mathcal H'_i$, has a finite outcome set
$\mathcal X$ and finite Kraus index sets $\mathcal R_x$. The physical outcome
is $x$ and $r:\mathcal R_x$ is an internal Kraus label:
\begin{equation}
\begin{gathered}
K_{x,r}:\mathcal H_i\longrightarrow\mathcal H'_i,
\qquad x:\mathcal X,\quad r:\mathcal R_x,\\
\sum_{x:\mathcal X}\sum_{r:\mathcal R_x}
 K_{x,r}^{\dagger}K_{x,r}=\openone_{\mathcal H_i}.
\end{gathered}
\end{equation}
\leanchor[https://maorbenshahar.github.io/QCryptLean/docs/QCryptLean/LOCC/Instrument.html\#LOCC.Instrument]{Instrument}

For a specified input system $\Gamma$, the structure PrivateAction
packages the acting party $i$, the outcome set $\mathcal X$ and the instrument $I$:
\begin{equation}
V=(i,\mathcal X,I).
\end{equation}
\leanchor[https://maorbenshahar.github.io/QCryptLean/docs/QCryptLean/LOCC/LocalAction.html\#LOCC.PrivateAction]{PrivateAction}
The output system $\Gamma'$ is computed. It is $\Gamma$ with party $i$'s register replaced by the output register of $I$, so $\Gamma'_j=\Gamma_j$ for every $j\ne i$ holds by construction.
Since $I$ acts on $\mathcal H_i$ alone, the action is local.

An announced action also specifies a report $g:\mathcal X\to\mathcal Y$,
where $\mathcal Y$ is a finite public alphabet and $g$ specifies which information about the outcome is announced. For a fixed-output instrument,
the constructor for the announced action takes the data
\begin{equation}
A=(i,\mathcal X,I,g).
\end{equation}
\leanchor[https://maorbenshahar.github.io/QCryptLean/docs/QCryptLean/LOCC/LocalAction.html\#LOCC.AnnouncedAction.ofInstrument]{ofInstrument}
The general AnnouncedAction structure also allows the acting party's output
register to depend on the report. It stores the Kraus family directly
\begin{equation}
K_{x,r}:\mathcal H_i\longrightarrow\mathcal H'_{i,g(x)},
\qquad y=g(x),
\end{equation}
\leanchor[https://maorbenshahar.github.io/QCryptLean/docs/QCryptLean/LOCC/LocalAction.html\#LOCC.AnnouncedAction]{AnnouncedAction}
with the same completeness condition as defined before. The output system of each public branch is computed in the same way.

Next, our goal is to construct some object that keeps track of private and announced actions and their dependent subsequent actions.
Write $A_j$ for Alice's
actions and $B_j$ for Bob's, schematically we want to represent trees such as
\[
A_1\left\{
\begin{array}{@{}l@{}}
\xrightarrow{a=0}B_1^{(0)}\to\mathsf{done},\\[1ex]
\xrightarrow{a=1}B_1^{(1)}
\left\{
\begin{array}{@{}l@{}}
\xrightarrow{b=0}A_2^{(0)}\to\mathsf{done},\\
\xrightarrow{b=1}A_2^{(1)}\to B_2\to\mathsf{done}.
\end{array}
\right.
\end{array}
\right.
\]
Here $A_1$ announces $a:\{0,1\}$, and $B_1^{(1)}$ announces
$b:\{0,1\}$. All remaining actions are private, and $\mathsf{done}$
marks termination. The three public histories are
$\mathcal T=\{(0),(1,0),(1,1)\}$.

In the library we introduce a type $\mathsf{Prog}(\Gamma)$ for holding onto the information in such a branching protocol, where $\Gamma$ is the system the protocol starts from. It is defined by the following three constructors
\begin{equation}
\begin{aligned}
&\mathsf{done},\\
&\mathsf{priv}(V,q), &&q:\mathsf{Prog}(\Gamma'(V)),\\
&\mathsf{announced}(A,y\mapsto q_y), &&q_y:\mathsf{Prog}(\Gamma'(A,y)).
\end{aligned}
\end{equation}
\leanchor[https://maorbenshahar.github.io/QCryptLean/docs/QCryptLean/LOCC/Program.html\#LOCC.Program]{Program}
Here $V$ and $A$ are the private and announced action records defined above, acting on $\Gamma$. Note that the systems $\Gamma'(V)$ and $\Gamma'(A,y)$ are not supplied by the author, they are determined by the actions as $\Gamma$ with the acting party's register replaced by the action's output register. A continuation must start from this system, otherwise the program does not type-check.

In practice the protocols are most easily assembled using a function $\mathsf{then}$ which continues an action with the program that follows it. For the example above, the \lean{} code would be
\begin{equation}
\label{eq:locc-sequencing}
\begin{aligned}
p={}&A_1.\mathsf{then}(\mathsf{fun}\ a\mapsto\mathsf{match}\ a\ \mathsf{with}\\
&\quad|\;0\Rightarrow B_1^{(0)}.\mathsf{then}(\mathsf{done})\\
&\quad|\;1\Rightarrow B_1^{(1)}.\mathsf{then}(\mathsf{fun}\ b\mapsto\mathsf{match}\ b\ \mathsf{with}\\
&\qquad\quad|\;0\Rightarrow A_2^{(0)}.\mathsf{then}(\mathsf{done})\\
&\qquad\quad|\;1\Rightarrow A_2^{(1)}.\mathsf{then}(B_2.\mathsf{then}(\mathsf{done})))).
\end{aligned}
\end{equation}
\leanchor[https://maorbenshahar.github.io/QCryptLean/docs/QCryptLean/LOCC/Program.html\#LOCC.AnnouncedAction.then]{AnnouncedAction.then}
\leanchor[https://maorbenshahar.github.io/QCryptLean/docs/QCryptLean/LOCC/Program.html\#LOCC.PrivateAction.then]{PrivateAction.then}

Programs can also be composed. For example, with the subprogram $q$ for the branch $a=1$, the same protocol reads
\begin{equation}
\begin{aligned}
q={}&B_1^{(1)}.\mathsf{then}(\mathsf{fun}\ b\mapsto\mathsf{match}\ b\ \mathsf{with}\\
&\quad|\;0\Rightarrow A_2^{(0)}.\mathsf{then}(\mathsf{done})\\
&\quad|\;1\Rightarrow A_2^{(1)}.\mathsf{then}(B_2.\mathsf{then}(\mathsf{done}))),\\
p={}&A_1.\mathsf{then}(\mathsf{fun}\ a\mapsto\mathsf{match}\ a\ \mathsf{with}\\
&\quad|\;0\Rightarrow B_1^{(0)}.\mathsf{then}(\mathsf{done})\\
&\quad|\;1\Rightarrow q).
\end{aligned}
\end{equation}

So far we have described how the library stores the information that defines a protocol in the $\mathsf{Prog}$ type. However, as it stands, a program $p:\mathsf{Prog}$ cannot be fed into a diamond norm or trace norm. To do that we still need to extract actual channels out of $p$, this process is described below.

Consider a particular branch in a program, with raw outcomes and internal Kraus indices fixed. We first extend each local Kraus map by the identity on the other parties' registers, then compose the resulting maps along the branch, including any private actions.
For the first two steps of the earlier example write $K_a^{\mathrm A}$ for Alice's Kraus map and $K_{b|a}^{\mathrm B}$ for Bob's (which may depend on the announcement $a$ that Alice makes). Taking one Kraus operator per outcome for simplicity, the channel for these two steps is
\begin{equation}
\begin{aligned}
K^{(2)}_{a,b}&=\bigl(\openone_{\mathrm A}\otimes K_{b|a}^{\mathrm B}\bigr)
          \bigl(K_a^{\mathrm A}\otimes\openone_{\mathrm B}\bigr),\\
\mathcal E^{(2)}(\rho)&=\sum_{a,b}K^{(2)}_{a,b}\rho{\bigl(K^{(2)}_{a,b}\bigr)^{\dagger}}
                  \otimes|t(a,b)\rangle\langle t(a,b)|.
\end{aligned}
\end{equation}
Here $t(0,b)=(0)$ and $t(1,b)=(1,b)$. Bob's outcome $b$ remains private when $a=0$ and is announced when $a=1$. Later steps multiply further local Kraus maps on the left of $K^{(2)}_{a,b}$.
The function $\mathsf{denote}$ does exactly this for every branch of the program and sums the resulting Kraus contributions to form the channel.
\leanchor[https://maorbenshahar.github.io/QCryptLean/docs/QCryptLean/LOCC/Program/Denotation.html\#LOCC.Program.denote]{denote}

\subsection{Security, robustness, and classicality}
\label{sec:protocol-properties}

Given a program $p:\mathsf{Prog}(\Gamma)$ its associated real channel is defined with the denote function we saw above,
\begin{equation}
\mathcal E^{\mathrm{real}}:=p.\mathsf{denote}.
\end{equation}
{\leanchor[https://maorbenshahar.github.io/QCryptLean/docs/QCryptLean/QKD/Protocol.html\#QKD.Protocol.real]{real}}
Since this channel only composes and sums the program's Kraus maps, it discards only the registers that those maps discard.
Therefore, registers for local operations whose outcomes are meant to be private must be explicitly traced out in the program before denotation.
If they are not, those registers remain in the channel output and enter the real--ideal security comparison as additional side information.
Tracing out cannot remove announced information, since it is a private action, and private actions never act on the public register.

To define security one compares the real channel to an ideal channel,
\begin{equation}
\mathcal E^{\mathrm{ideal}}:=\mathcal R\circ\mathcal E^{\mathrm{real}},
\end{equation}
{\leanchor[https://maorbenshahar.github.io/QCryptLean/docs/QCryptLean/QKD/Protocol.html\#QKD.Protocol.ideal]{ideal}}
where $\mathcal R$ is a channel that erases the keys that Alice and Bob produced and writes
one fresh, uniform random string of the same length into both key registers.
It preserves the public transcript and all other registers, and leaves
aborted runs unchanged. Therefore, on each accepting branch the ideal keys are equal, uniform, and independent of Eve's information.
Each $\mathsf{done}$ in a $\mathsf{Prog}$ can also carry an optional label; in a QKD program, $\mathcal R$ uses these labels to determine which registers are the keys.
\leanchor[https://maorbenshahar.github.io/QCryptLean/docs/QCryptLean/QKD/KeyEnd.html\#QKD.KeyEnd]{KeyEnd}

For a QKD protocol with classical output keys, $\varepsilon$-security is
\begin{equation}
\label{eq:protocol-security}
\tfrac12\|\mathcal E^{\mathrm{real}}-\mathcal E^{\mathrm{ideal}}\|_\diamond
 \le\varepsilon,
\end{equation}
\leanchor[https://maorbenshahar.github.io/QCryptLean/docs/QCryptLean/QKD/Protocol.html\#QKD.Protocol.IsSecure]{IsSecure}
where the diamond norm is
\begin{equation}
\|\Lambda\|_\diamond=
\sup_{\|X\|_1\le1}\|(\Lambda\otimes\mathrm{id}_E)(X)\|_1,
\end{equation}
\leanchor[https://maorbenshahar.github.io/QCryptLean/docs/QCryptLean/Quantum/Channels/Diamond.html\#Quantum.Channels.diamondNorm]{diamondNorm}
and $\|X\|_1=\Tr\sqrt{X^\dagger X}$ with $X$ ranging over operators on the input together with a reference $E$ of the same dimension. Half the diamond norm is the distinguishing advantage~\cite{PortmannRenner21}, so the real and ideal outputs have trace distance at most $\varepsilon$.
Applying the same subsequent channel to both outputs
cannot increase their distinguishability, including processing
conditioned on the public transcript and acting jointly with Eve's reference.
\leanchor[https://maorbenshahar.github.io/QCryptLean/docs/QCryptLean/Quantum/Channels/DiamondInequality.html\#Quantum.Channels.diamondDist_comp_le]{diamondDist\_comp\_le}

According to this definition, a protocol that always aborts will appear secure, hence this definition of security is coupled with robustness which bounds the abort probability. Summing over the aborting public histories for a given input state $\rho$ gives
\begin{equation}
p_{\mathrm{abort}}(\rho)
 :=\sum_{t\,\mathrm{aborts}}\Tr\!\left([\mathcal E^{\mathrm{real}}(\rho)]_t\right),
\end{equation}
{\leanchor[https://maorbenshahar.github.io/QCryptLean/docs/QCryptLean/QKD/Acceptance.html\#QKD.Protocol.abortProbability]{abortProbability}}
where $t$ selects the output block for that history. The robustness bound for a specified input state, or class of inputs, is
\begin{equation}
\label{eq:protocol-robustness}
p_{\mathrm{abort}}(\rho)\le\delta_{\mathrm{rob}},
\qquad 0\le\delta_{\mathrm{rob}}<1.
\end{equation}

We also define classicality of Alice's and Bob's actions, which Sec.~\ref{sec:bb84-protocol} uses for the BB84 post-processing. A program is classical when every
local action preserves diagonality on Alice's and Bob's registers for
each outcome, along every branch. The library checks this property recursively.
\leanchor[https://maorbenshahar.github.io/QCryptLean/docs/QCryptLean/LOCC/Program/TwoPartyClassicalStorage.html\#LOCC.Program.IsHonestClassical]{IsHonestClassical}
For a joint input that is classical on Alice's and Bob's registers and quantum
on the reference, such a program has an equivalent implementation by
local stochastic operations and classical communication.
\leanchor[https://maorbenshahar.github.io/QCryptLean/docs/QCryptLean/LOCC/Program/TwoPartyClassicalReplacement.html\#LOCC.Program.IsHonestClassical.tensorIdLinear_denote_classicalReplacement]{tensorIdLinear\_denote\_classicalReplacement}

\section{Machine-verified security of BB84}
\label{sec:bb84-realization}

\begin{figure}[tb]
\centering
\includegraphics[width=0.9\columnwidth]{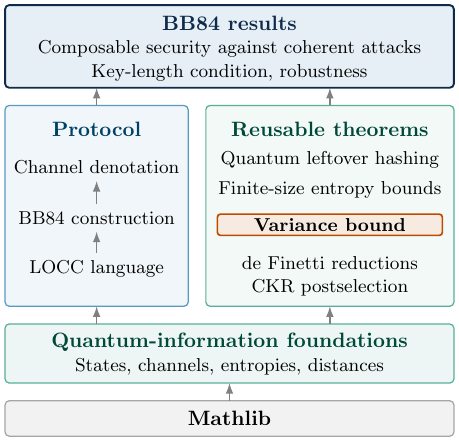}
\caption{\textbf{The machine-checked BB84 development} (read bottom-up). The LOCC protocol construction and reusable quantum-information theorems support the BB84 security, key-length, and robustness results. The highlighted box marks the improved conditional-entropy variance bound of Sec.~\ref{sec:discovery}.}
\label{fig:stack}
\end{figure}

We now apply the definitions of Sec.~\ref{sec:lean-locc} to entanglement-based qubit BB84. We construct the protocol, state its finite-key security theorem, and bound its abort probability on honest inputs. The \lean{} kernel checks the theorems of this section and of Sec.~\ref{sec:discovery}, and their proofs (Fig.~\ref{fig:stack}) use only \lean{}'s standard foundational axioms (\texttt{propext}, \texttt{Classical.choice} and \texttt{Quot.sound}).

\subsection{The protocol}
\label{sec:bb84-protocol}

The protocol is a program in the LOCC language of Sec.~\ref{sec:lean-locc}. Alice and Bob share an authenticated public classical channel, and initially hold $N$ qubit pairs possibly distributed by Eve who may keep a purification.

Alice and Bob measure every qubit in either the $Z$ or the $X$ basis, drawing the basis choice randomly from their respective distributions $p_A$ and $p_B$. A party with basis distribution $p$ measures with the instrument
\begin{equation}
\label{eq:bb84-kraus}
K^p_{\theta,x}=\sqrt{p(\theta)}\,|\theta,x\rangle\langle\phi^\theta_x|,
\end{equation}
where $\theta:\{Z,X\}$ is the basis, $|\phi^\theta_x\rangle$ is the eigenstate of the Pauli operator $\theta$ with eigenvalue $(-1)^x$, and $|\theta,x\rangle$ is an orthogonal classical record of basis and outcome. The qubit is discarded and nothing is announced.
\leanchor[https://maorbenshahar.github.io/QCryptLean/docs/QCryptLean/QKD/BB84/Measurement/Weighted.html\#QKD.BB84.Measurement.weightedMeasureAndRecord]{weightedMeasureAndRecord}

This is followed by classical post-processing. First Alice and Bob announce their basis strings, and Alice announces a uniformly random permutation of the rounds with agreeing bases.
In the order of this permuted register, the first $n_K$ matched $Z$ rounds form the raw key, the next $m_Z$ the $Z$ test block and the first $m_X$ matched $X$ rounds the $X$ test block, and each party keeps only its outcome bits in these
\begin{equation}
\label{eq:retained-block}
n:=n_K+m_Z+m_X
\end{equation}
retained rounds; if the matched rounds cannot fill the three blocks, both parties instead discard their records and abort.

This is followed by parameter estimation, error correction and verification, and privacy amplification, which are not specific to BB84.
In parameter estimation, Bob and then Alice announce their bits in every test round.

Alice then draws a verification seed and a privacy-amplification seed, binary matrices with $n_K$ columns and with $\ell_{\mathrm{EV}}$ and $\ell$ rows, uniformly and independently.
She announces both seeds, the $\ell_{\mathrm{EV}}$-bit verification tag given by the product of the verification seed and her raw key $x$ modulo~2, and the $\ell_{\mathrm{EC}}$-bit syndrome $S(x)$ of an error-correction scheme with syndrome map $S$ and decoder $D$.

Bob computes his corrected key $D(y,S(x))$ from his raw key $y$.
He announces acceptance if each test block is nonempty and its fraction of disagreeing bits lies in the window $[Q-\delta,Q+\delta]$, with center $Q$ and half-width $\delta$, and if the tag of his corrected key equals Alice's; otherwise he announces abort.

On acceptance, Alice's final key is the $\ell$-bit product of the privacy-amplification seed and her raw key modulo~2, and Bob's is the same product with his corrected key.
The parties discard all other private records on acceptance and all their private records on abort.
\leanchor[https://maorbenshahar.github.io/QCryptLean/docs/QCryptLean/QKD/BB84/Program.html\#QKD.BB84.construction]{construction}

The measurement step leaves classical records by construction, and every action of the post-processing preserves diagonality on Alice's and Bob's registers, so the post-processing is classical in the sense of Sec.~\ref{sec:protocol-properties}.
\leanchor[https://maorbenshahar.github.io/QCryptLean/docs/QCryptLean/QKD/BB84/ClassicalStages.html\#QKD.BB84.exists_eq_measureRounds_isHonestClassical]{exists\_eq\_measureRounds\_isHonestClassical}

\subsection{Finite-key security theorem}
\label{sec:bb84-theorem}

We state the bounds first, then explain their terms.

\textbf{Theorem 1 (finite-key security of BB84).}
For every translation-equivariant error-correction scheme (Appendix~\ref{app:decoder}) and every choice of analysis parameters satisfying the conditions below,
the real and ideal channels of the protocol satisfy
\begin{equation}
\label{eq:typed-bb84-security}
\tfrac12\|\mathcal E^{\mathrm{real}}_{\mathrm{BB84}}-\mathcal E^{\mathrm{ideal}}_{\mathrm{BB84}}\|_\diamond\le\epsqkd,
\end{equation}
so the protocol is $\epsqkd$-secure in the sense of Eq.~\eqref{eq:protocol-security}, where $\epsqkd$ is the error-verification error plus the postselection prefactor $g_{n,4}$ times the sum of the contributions of parameter estimation, smoothing and privacy amplification:
\begin{equation}
\label{eq:typed-bb84-budget}
\begin{gathered}
\epsqkd:=\epsEV+{}g_{n,4}\bigl(\epsPA+2\bar\varepsilon+2\sqrt{2\epsAT}\bigr).
\end{gathered}
\end{equation}
\leanchor[https://maorbenshahar.github.io/QCryptLean/docs/QCryptLean/QKD/BB84/Security/BellRenyi.html\#QKD.BB84.Parameters.bellRenyiBudget]{bellRenyiBudget}
Here $g_{n,4}:=\binom{n+3}{3}$ is the postselection prefactor for extending the IID analysis to arbitrary inputs~\cite{CKR09,Nahar24}, and the bracket matches the bound of Ref.~\cite[Theorem~3]{Nahar24} for mixtures of IID inputs. The prefactor is the dimension of the symmetric subspace of $(\mathbb C^4)^{\otimes n}$. Generic postselection for two qubits per round gives the larger prefactor $\binom{n+15}{15}$, from the symmetric subspace of $(\mathbb C^{16})^{\otimes n}$. The smaller factor exploits BB84's Pauli symmetry: for a translation-equivariant error-correction scheme, applying the same Pauli operator to both qubits of any round relabels the classical outputs~\cite[Lemma~2]{Nahar24}. We use the acceptance-test and verification bounds
\begin{equation}
\label{eq:typed-bb84-errors}
\begin{aligned}
\epsAT &:=\exp\bigl[-m_X\,\mathrm{kl}(Q+\delta\,\|\,Q+\delta+\delta')\bigr],\\
\epsEV&:=2^{-\ell_{\mathrm{EV}}},
\end{aligned}
\end{equation}
\leanchor[https://maorbenshahar.github.io/QCryptLean/docs/QCryptLean/QKD/BB84/FiniteKey/ParameterEstimation/PhaseTail.html\#QKD.BB84.FiniteKey.phaseTail]{phaseTail}
where $\mathrm{kl}(a\|b):=a\ln(a/b)+(1-a)\ln[(1-a)/(1-b)]$ is the binary relative entropy in nats. Here $\epsAT$ bounds the probability that an IID input whose $X$-basis error rate exceeds the phase-error bound $Q+\delta+\delta'$ passes parameter estimation, by the Chernoff bound for the $X$ test~\cite{Mulzer18}. $\epsEV$ bounds the probability that a corrected key differing from Alice's raw key passes verification for a uniformly random verification seed. This holds for every input, so $\epsEV$ is not multiplied by $g_{n,4}$. The deviation $\delta'$, the smoothing parameter $\bar\varepsilon$ and the privacy-amplification error $\epsPA$ are chosen in the analysis. 
\leanchor[https://maorbenshahar.github.io/QCryptLean/docs/QCryptLean/QKD/BB84/FiniteKey/ParameterEstimation/PhasePivot.html\#QKD.BB84.FiniteKey.componentAcceptProbability_le_exp_neg_mul_klBer]{componentAcceptProbability\_le\_exp\_neg\_mul\_klBer}

The bound holds provided that $\bar\varepsilon$, $\epsPA$, $\beta$ and $\delta'$ are positive, that
\begin{equation}
\label{eq:typed-bb84-regime}
\begin{gathered}
Q+\delta+\delta'\le\tfrac12,\qquad\beta<1,
\end{gathered}
\end{equation}
and that the final key length obeys
\begin{equation}
\label{eq:typed-bb84-key-length}
\begin{aligned}
\ell\le{}&n_K\bigl[1-h_2(Q+\delta+\delta')\bigr]-\ell_{\mathrm{EC}}-\ell_{\mathrm{EV}}\\
&-2\log_2g_{n,4}-P_\beta(n_K,\bar\varepsilon)-2\log_2(1/\epsPA){}+2,
\end{aligned}
\end{equation}
\leanchor[https://maorbenshahar.github.io/QCryptLean/docs/QCryptLean/QKD/BB84/Security/BellRenyi.html\#QKD.BB84.Parameters.isSecure_of_bellRenyiConditions]{isSecure\_of\_bellRenyiConditions}

The key-length condition~\eqref{eq:typed-bb84-key-length} starts from the entropy of the raw key at the phase-error bound, with $h_2$ the binary entropy in bits, and subtracts, in turn, the public leakage of the syndrome and the tag, the postselection charge, the finite-size penalty and the hashing cost of the leftover-hash lemma~\cite{RennerKonig05,Renner08}, as required by Ref.~\cite[Eq.~(11)]{Nahar24}. The $+2$ arises because $\epsPA$ bounds a trace distance, half of a trace norm. The finite-size penalty
\begin{equation}
\label{eq:coriv2penalty}
\begin{aligned}
P_\beta(n_K,\bar\varepsilon)&:=\frac{C_\star\ln2}{2}\,n_K\beta+K_\beta\,n_K\beta^2+\frac{b(\bar\varepsilon)}{\beta},\\
b(\bar\varepsilon)&:=\log_2(2/\bar\varepsilon^2),\qquad C_\star:=\log_2^2\bigl(1+\sqrt2\bigr),\\
K_\beta&{}:=\frac{K_\star\,2^\beta}{(1-\beta)^3},\qquad K_\star:=\frac{\ln^3(2+e^2)}{6\ln2},
\end{aligned}
\end{equation}
\leanchor[https://maorbenshahar.github.io/QCryptLean/docs/QCryptLean/InfoTheory/Renyi/FiniteSizePenalty.html\#InfoTheory.Renyi.renyiPenalty]{renyiPenalty}
bounds how far the smooth min-entropy of the raw key can fall below $n_K$ times its per-round von Neumann entropy. Section~\ref{sec:coriv2math} shows where each term comes from, the first through the variance cap $C_\star$. 

The parameters $\delta'$, $\bar\varepsilon$, $\epsPA$ and $\beta$ are free and trade the error against the key length. A larger deviation $\delta'$ between the window edge $Q+\delta$ and the phase-error bound lowers $\epsAT$ but raises $h_2(Q+\delta+\delta')$. The condition $Q+\delta+\delta'\le\frac12$ keeps $h_2$ increasing up to the bound, so every input below it has entropy at least $1-h_2(Q+\delta+\delta')$ per round. Smaller $\bar\varepsilon$ and $\epsPA$ lower the error but raise $P_\beta$ and $2\log_2(1/\epsPA)$, and $\beta$ balances the terms of $P_\beta$ within the range $0<\beta<1$ of Ref.~\cite[Cor.~IV.2]{DupuisFawzi19}.

\subsection{Robustness}
\label{sec:bb84-robustness}

Theorem~2 bounds the abort probability of Sec.~\ref{sec:protocol-properties} on an honest channel, which is the robustness of Ref.~\cite{PortmannRenner21}.

\textbf{Theorem 2 (robustness of BB84).}
Let a memoryless channel without an eavesdropper deliver $\sigma^{\otimes N}$, where in matching bases the two-qubit state $\sigma$ gives uniform outcome bits that disagree with probability $q$, the quantum bit error rate (QBER). Let Bob's corrected key differ from Alice's raw key with probability at most $\varepsilon_{\mathrm{EC}}$, for every raw key of Alice's, when his raw key differs from hers by independent bit flips with probability $q$. Let the quotas $k_Z:=n_K+m_Z$ and $k_X:=m_X$ of matched rounds required by selection, the matched-basis probabilities $P_\theta:=p_A(\theta)p_B(\theta)$ and the window satisfy
\begin{equation}
\label{eq:quota-condition}
\begin{gathered}
\frac{k_\theta-1}{N}<P_\theta<1\quad(\theta:\{Z,X\}),\\
0<Q-\delta<q<Q+\delta<1.
\end{gathered}
\end{equation}
Then
\begin{equation}
\label{eq:honest-abort}
p_{\mathrm{abort}}(\sigma^{\otimes N})\le\varepsilon_{\mathrm{quota}}+\varepsilon_{\mathrm{test}}+\varepsilon_{\mathrm{EC}},
\end{equation}
where the relative-entropy Chernoff bounds~\cite[Thm.~2.1 and Cor.~4.1]{Mulzer18}
\begin{equation}
\label{eq:abort-terms}
\begin{aligned}
\varepsilon_{\mathrm{quota}}&:=\sum_{\theta:\{Z,X\}}e^{-N\mathrm{kl}((k_\theta-1)/N\,\|\,P_\theta)},\\
\varepsilon_{\mathrm{test}}&:=\sum_{\theta:\{Z,X\}}\bigl[e^{-m_\theta\mathrm{kl}(Q+\delta\|q)}+e^{-m_\theta\mathrm{kl}(Q-\delta\|q)}\bigr]
\end{aligned}
\end{equation}
bound the probabilities that a quota is not met and that a test block leaves the window.
\leanchor[https://maorbenshahar.github.io/QCryptLean/docs/QCryptLean/QKD/BB84/QuantitativeRobustness.html\#QKD.BB84.Parameters.one_sub_abortBound_le_acceptProbability]{one\_sub\_abortBound\_le\_acceptProbability}

Proposition~4 in Appendix~\ref{app:decoder} gives a translation-equivariant error-correction scheme with this property, a linear syndrome of slightly more than $n_Kh_2(q)$ bits, so Theorems~1 and~2 hold for the same protocol.

\section{A tighter finite-key bound from the pipeline}
\label{sec:discovery}

This section describes the origin of the key length of Theorem~1. Its finite-size penalty rests on a new inequality, found with the pipeline of Sec.~\ref{sec:method}: a bound on the conditional-entropy variance of classical--quantum states that improves that of Ref.~\cite{DupuisFawzi19} (Sec.~\ref{sec:coriv2math}).

\subsection{Variance bound and finite-size penalty}
\label{sec:coriv2math}

Let $\rho_{WB}$ be a normalized classical--quantum state, with $W$ of dimension $d_{W}$ and arbitrary finite-dimensional quantum side information $B$. Its conditional-entropy variance~\cite[Defs.~III.1--III.2]{DupuisFawzi19}, in bits squared, is:
\begin{equation}
\label{eq:conditional-entropy-variance}
\begin{aligned}
L_\rho&:=\log_2\rho_{WB}-\log_2(\openone_{W}\otimes\rho_B),\\
V(W|B)_\rho&:=\Tr(\rho_{WB}L_\rho^2)-\bigl[\Tr(\rho_{WB}L_\rho)\bigr]^2,
\end{aligned}
\end{equation}
where $\rho_B$ is the marginal on $B$ and each logarithm is taken on its support.

\textbf{Theorem 3 (conditional-entropy variance).}
Every such state satisfies
\begin{equation}
\label{eq:variance-general-improved}
V(W|B)_\rho\le\log_2^2\bigl(\sqrt{d_{W}}+\sqrt{d_{W}-1}\bigr).
\end{equation}
\leanchor[https://maorbenshahar.github.io/QCryptLean/docs/QCryptLean/InfoTheory/Renyi/Bounds.html\#InfoTheory.Renyi.condVariance_le_logb_sq]{condVariance\_le\_logb\_sq}
This improves the bound $\log_2^2(2d_{W}+1)$ of Ref.~\cite[Cor.~III.5]{DupuisFawzi19} for every $d_{W}$. For a binary key it gives the cap used by Theorem~1,
\begin{equation}
\label{eq:coriv2caps}
V(W|B)_\rho\le C_\star=\log_2^2\bigl(1+\sqrt2\bigr)\approx1.6168,
\end{equation}
\leanchor[https://maorbenshahar.github.io/QCryptLean/docs/QCryptLean/InfoTheory/Renyi/QubitVariance.html\#InfoTheory.Renyi.condVariance_le_logb_sq_of_card_eq_two]{condVariance\_le\_logb\_sq\_of\_card\_eq\_two}
in place of $\log_2^25\approx5.39$, both in bits squared.

\emph{Proof sketch.} Write $\rho_{WB}=\sum_x\lambda_x|e_x\rangle\langle e_x|$ and $\openone_{W}\otimes\rho_B=\sum_y\kappa_y|f_y\rangle\langle f_y|$ in eigenbases. The Nussbaum--Szko\l{}a weights $\mu(x,y):=\lambda_x|\langle e_x|f_y\rangle|^2$ and $\nu(x,y):=\kappa_y|\langle e_x|f_y\rangle|^2$~\cite{DupuisFawzi19} are a probability distribution and a nonnegative weight function with $\sum\nu=\Tr(\openone_{W}\otimes\rho_B)=d_{W}$, and $V(W|B)_\rho\ln^22$ is the variance under $\mu$ of the log-likelihood ratio $Y:=\ln(\mu/\nu)$. It satisfies $\mathbb E[e^{Y}]=\Tr\rho_{WB}^2(\openone_{W}\otimes\rho_B)^{-1}\le1$, with the inverse on the support, because $\rho_{WB}\le\openone_{W}\otimes\rho_B$. It also satisfies $\mathbb E[e^{-Y}]=\sum_{\mu>0}\nu\le d_{W}$. With the shift $c:=-\frac12\ln d_{W}$ these bounds give $\mathbb E[\cosh(Y-c)]\le\sqrt{d_{W}}$. Let $r:=\ln\bigl(\sqrt{d_{W}}+\sqrt{d_{W}-1}\bigr)$, so that $\cosh r=\sqrt{d_{W}}$. Every real $z$ satisfies $z^2\le r^2+\frac{2r}{\sinh r}(\cosh z-\cosh r)$, where $2r/\sinh r$ is read as $2$ when $d_{W}=1$. Since the mean minimizes the mean squared deviation, taking $z=Y-c$ and averaging under $\mu$ gives
\begin{equation*}
\begin{aligned}
\operatorname{Var}Y&\le\mathbb E[(Y-c)^2]\\
&\le r^2+\frac{2r}{\sinh r}\bigl(\mathbb E[\cosh(Y-c)]-\cosh r\bigr)\le r^2,
\end{aligned}
\end{equation*}
where the last step uses $\mathbb E[\cosh(Y-c)]\le\sqrt{d_{W}}=\cosh r$. Dividing by $\ln^22$ gives~\eqref{eq:variance-general-improved}.

For a binary key and $n_K$ independent rounds, the finite-size penalty bounds the gap between $n_K H(W|B)_\rho$, with $H(W|B)_\rho:=-\Tr(\rho_{WB}L_\rho)$, and the smooth min-entropy that the leftover-hash lemma turns into key: for every $0<\beta<1$,

\begin{equation*}
n_K H(W|B)_\rho-H_{\min}^{\bar\varepsilon}(W^{n_K}|B^{n_K})\le P_\beta(n_K,\bar\varepsilon).
\end{equation*}
\leanchor[https://maorbenshahar.github.io/QCryptLean/docs/QCryptLean/QKD/BB84/FiniteKey/PerRound/RenyiPenalty.html\#QKD.BB84.FiniteKey.le_smoothMinEntropy_tensorPower_renyiPenalty]{le\_smoothMinEntropy\_tensorPower\_renyiPenalty}
To see where the variance enters, split the gap at the Petz conditional R\'enyi entropy $H'_{1+\beta}(W|B)_\rho:=-\frac1\beta\log_2\Tr\rho_{WB}^{1+\beta}(\openone_{W}\otimes\rho_B)^{-\beta}$, with the negative power on the support:
\begin{equation*}
\begin{aligned}
&n_K H-H_{\min}^{\bar\varepsilon}=\bigl[n_K H'_{1+\beta}-H_{\min}^{\bar\varepsilon}\bigr]+n_K\bigl[H-H'_{1+\beta}\bigr]\\
&\quad\le\frac{b(\bar\varepsilon)}{\beta}+n_K\Bigl[\frac{\beta\ln2}{2}V+\beta^2K(1+\beta)\Bigr]\\
&\quad\le\frac{b(\bar\varepsilon)}{\beta}+n_K\Bigl[\frac{C_\star\ln2}{2}\beta+K_\beta\beta^2\Bigr]=P_\beta,
\end{aligned}
\end{equation*}
with the arguments dropped. The first bracket is bounded by the conversion from R\'enyi to smooth min-entropy~\cite[Prop.~6.5 and Eqs.~(6.91)--(6.92)]{Tomamichel16}, and the second by the continuity bound of Ref.~\cite[Lemma~IV.1]{DupuisFawzi19}, with $K(1+\beta)$ given in Ref.~\cite[Cor.~IV.2]{DupuisFawzi19}. Theorem~3 then caps $V$ by $C_\star$, and $K(1+\beta)\le K_\beta$.

\section{Discussion and Outlook}
\label{sec:disc}

We used an AI-assisted proof pipeline to connect a typed LOCC description of BB84 to an end-to-end composable-security theorem. With the same pipeline we found a tighter bound on the conditional-entropy variance, which reduces the finite-size penalty in the key length of Theorem~1 (Sec.~\ref{sec:discovery}).

The formalization also made hidden assumptions in the literature explicit. Let us give two examples, the proof of the postselection theorem of Ref.~\cite{CKR09} restricts the diamond norm to density-operator inputs, which requires the map to preserve Hermiticity. This condition holds for the differences of channels to which the theorem is applied, and the formal statement assumes it. Likewise, the continuity bound of Ref.~\cite[Lemma~IV.1]{DupuisFawzi19} holds when the support of $\rho$ lies in that of $\sigma$, a condition its statement leaves implicit and the formal statement adds.

Several components of this development can also be used independently of BB84. These include quantum leftover hashing~\cite{RennerKonig05,Renner08} with its public seed retained in the security comparison, smooth-entropy bounds for classical--quantum tensor-power states derived from conditional R\'enyi entropy, and the postselection reduction of Ref.~\cite{CKR09}.
{\leanchor[https://maorbenshahar.github.io/QCryptLean/docs/QCryptLean/InfoTheory/QuantumLHL/Main.html\#InfoTheory.QuantumLHL.SeedKey.traceDistanceGen_output_le_of_le_smoothMinEntropyOpt]{traceDistanceGen\_output\_le\_of\_le\_smoothMinEntropyOpt}}
A new protocol would supply its own program and parameter estimation and reuse these results.
Since the protocol language allows any finite set of parties, it could also describe protocols beyond key distribution, such as secure computation and digital signatures~\cite{BozzioBeyondQKD25} or multiparty deep learning~\cite{SulimanyMPDL25}, each with its own security definition.

The library can be extended in several directions.
To represent prepare-and-measure protocols directly, the protocol language could be extended to describe quantum transmission followed by LOCC processing. A more specialized protocol type could also enforce the measurement and subsequent classical-processing stages by construction, while security and robustness would remain properties to prove.
Formal security proofs of optical QKD with decoy states~\cite{Hwang03,LoMaChen05,MaQiZhaoLo05} would additionally require suitable source, loss, and detector models.
The same protocol definitions could support security proofs based on entropy accumulation~\cite{DupuisEA20} or uncertainty relations~\cite{TomamichelFiniteKey12}, allowing their assumptions and finite-size guarantees to be compared.

Several of the formal results above were written by the open-weight GLM-5.3-Flash model of Sec.~\ref{sec:method}, run locally through \texttt{loopy} under the direction of Claude Opus 5. GLM worked on the generalization of Theorem~1 to a free privacy-amplification error, R\'enyi offset and deviation, and the relative-entropy tails of Theorem~2 and Proposition~4. The frontier model edited loopy prompts to tune the tasks of GLM, and Claude and GPT reviewed its PRs before they were merged into the library. GLM had \lean{} language-server tools but no shell. We hope to develop theorem proving with open-weight models further, through well-tuned harnesses and reinforcement learning.

\begin{acknowledgments}
We wish to thank Shlok Nahar, Devashish Tupkary, Uri Kol, Winston Yin, and Benjy Firester for useful discussions.

K.S.\ acknowledges support from the European Union's Horizon Europe research and innovation programme under the Marie Sk\l{}odowska-Curie grant agreement No.~101202109. The research of M.B.S.\ is supported by the Knut and Alice Wallenberg Foundation, Grant KAW~2023.0490. J.M.\ acknowledges support from the MITRE Quantum Moonshot program, and the MIT Kalaniyot fellowship.

\end{acknowledgments}

\appendix
\renewcommand{\thesubsection}{\thesection.\arabic{subsection}}

\section{An admissible decoder}
\label{app:decoder}

An error-correction scheme with syndrome map $S$ and decoder $D$ is translation-equivariant if for every bit string $u$ there is a permutation $\pi_u$ of the syndrome values with $S(x\oplus u)=\pi_u\bigl(S(x)\bigr)$ and $D\bigl(y\oplus u,S(x\oplus u)\bigr)=D\bigl(y,S(x)\bigr)\oplus u$ for all raw keys $x$ and $y$, where $\oplus$ is bitwise addition modulo~2.
\leanchor[https://maorbenshahar.github.io/QCryptLean/docs/QCryptLean/QKD/BB84/ErrorCorrection.html\#QKD.BB84.ECScheme.IsTranslationEquivariant]{IsTranslationEquivariant}

\textbf{Proposition 4 (an admissible decoder).}
Let $q>0$, and let a decoding radius $t$ and a number $s$ of extra syndrome bits, both natural numbers, and a relative radius $r$ satisfy
\begin{equation}
\label{eq:decoder-conditions}
\begin{gathered}
0\le r\le\tfrac12,\qquad t\le rn_K,\qquad qn_K<t+1\le n_K,\\
n_Kh_2(r)+s\le\ell_{\mathrm{EC}}.
\end{gathered}
\end{equation}
Then there is a translation-equivariant error-correction scheme with an $\ell_{\mathrm{EC}}$-bit syndrome that meets the decoding hypothesis of Theorem~2 at QBER $q$ with
\begin{equation}
\label{eq:decoder-failure}
\varepsilon_{\mathrm{EC}}=e^{-n_K\mathrm{kl}((t+1)/n_K\,\|\,q)}+2^{-s}.
\end{equation}
\leanchor[https://maorbenshahar.github.io/QCryptLean/docs/QCryptLean/QKD/BB84/LinearReconciliation.html\#QKD.BB84.ECScheme.exists_decodesWhp_of_entropy_le]{exists\_decodesWhp\_of\_entropy\_le}

The two terms bound the probabilities that more than $t$ errors occur~\cite[Thm.~2.1]{Mulzer18} and that another error pattern of weight at most $t$ has the same syndrome. The scheme is a linear syndrome with coset decoding, shown to exist by averaging over all binary linear syndrome maps; the proposition provides neither an explicit code nor an efficient decoder.

\section{Glossary of formal definitions}
\label{sec:glossary}

Table~\ref{tab:glossary} collects the main formal objects of the development and the \lean{}
identifiers that realize them. It lists definitions only, grouped into the standard
quantum-information primitives, the LOCC program language of Sec.~\ref{sec:lean-locc}, and the
security notions built on them. The identifiers match the names used in the text. An audit of the dependencies of the theorems of Secs.~\ref{sec:bb84-realization} and~\ref{sec:discovery} finds only the standard axioms \texttt{propext}, \texttt{Classical.choice} and \texttt{Quot.sound}.

\begin{table*}[t]
\caption{\textbf{Glossary of the main formal definitions.} Symbol or term,
the \lean{} identifier that realizes it, and its meaning. $\mathcal H$ is a party's finite register, $\rho,\sigma$
are states, $\Lambda$ a linear map, and $E$ a reference system.}
\label{tab:glossary}
\begin{ruledtabular}
\footnotesize
\renewcommand{\arraystretch}{1.18}
\begin{tabular}{l p{0.25\textwidth} p{0.50\textwidth}}
\textbf{Symbol / term} & \textbf{\lean{} identifier} & \textbf{Meaning}\\
\colrule
\multicolumn{3}{l}{\textit{Operators and channels (standard quantum information)}}\\[3pt]
$\mathrm{Op}(\mathcal H)$ & \leanlink[https://maorbenshahar.github.io/QCryptLean/docs/QCryptLean/Quantum/Operators/Basic.html\#Quantum.Operators.Op]{Op} & Operators on a party's finite register; the complex matrices on~$\mathcal H$.\\
$\Lambda$ & \leanlink[https://maorbenshahar.github.io/QCryptLean/docs/QCryptLean/Quantum/Channels/Basic.html\#Quantum.Channels.IsChannel]{IsChannel} & Predicate that a linear map is completely positive and trace preserving, i.e.\ a quantum channel.\\
$\|\cdot\|_\diamond$ & \leanlink[https://maorbenshahar.github.io/QCryptLean/docs/QCryptLean/Quantum/Channels/Diamond.html\#Quantum.Channels.diamondNorm]{diamondNorm} & Diamond norm $\sup_{\|X\|_1\le1}\|(\Lambda\otimes\mathrm{id}_E)X\|_1$ over a same-dimension reference $E$; half the diamond norm of the difference of two channels is the largest trace distance between their outputs.\\
$D_{\mathrm{tr}}(\rho,\sigma)$ & \leanlink[https://maorbenshahar.github.io/QCryptLean/docs/QCryptLean/Quantum/Metrics/Basic.html\#Quantum.Metrics.traceDistance]{traceDistance} & Trace distance $\tfrac12\|\rho-\sigma\|_1$ between states.\\
$F(\rho,\sigma)$ & \leanlink[https://maorbenshahar.github.io/QCryptLean/docs/QCryptLean/Quantum/Metrics/Basic.html\#Quantum.Metrics.fidelity]{fidelity} & Fidelity $\|\sqrt{\rho}\sqrt{\sigma}\|_1$ between states (squared form \leanlink[https://maorbenshahar.github.io/QCryptLean/docs/QCryptLean/Quantum/Metrics/Basic.html\#Quantum.Metrics.fidelitySq]{fidelitySq}).\\
\addlinespace[3pt]
\colrule
\multicolumn{3}{l}{\textit{LOCC program language}}\\[3pt]
$\Gamma$ & \leanlink[https://maorbenshahar.github.io/QCryptLean/docs/QCryptLean/LOCC/MultipartiteSystem.html\#LOCC.MultipartiteSystem]{MultipartiteSystem} & Local register bases of a multipartite quantum system.\\
instrument & \leanlink[https://maorbenshahar.github.io/QCryptLean/docs/QCryptLean/LOCC/Instrument.html\#LOCC.Instrument]{Instrument} & Quantum instrument on one party's register~\cite[Sec.~II]{ChitambarLOCC14}.\\
private action & \leanlink[https://maorbenshahar.github.io/QCryptLean/docs/QCryptLean/LOCC/LocalAction.html\#LOCC.PrivateAction]{PrivateAction} & Local action whose classical outcome is kept private.\\
announced action & \leanlink[https://maorbenshahar.github.io/QCryptLean/docs/QCryptLean/LOCC/LocalAction.html\#LOCC.AnnouncedAction]{AnnouncedAction} & Local action that publicly announces a coarse-graining of its outcome.\\
then & \leanlink[https://maorbenshahar.github.io/QCryptLean/docs/QCryptLean/LOCC/Program.html\#LOCC.PrivateAction.then]{PrivateAction.then} & Sequences an action with its continuation (announced analogue \leanlink[https://maorbenshahar.github.io/QCryptLean/docs/QCryptLean/LOCC/Program.html\#LOCC.AnnouncedAction.then]{AnnouncedAction.then}).\\
$p:\mathsf{Prog}(\Gamma)$ & \leanlink[https://maorbenshahar.github.io/QCryptLean/docs/QCryptLean/LOCC/Program.html\#LOCC.Program]{Program} & An LOCC protocol as a tree of private and announced actions, constructible only by composing local actions.\\
$\mathcal{B}$ & \leanlink[https://maorbenshahar.github.io/QCryptLean/docs/QCryptLean/LOCC/Boundary.html\#LOCC.Boundary]{Boundary} & The public transcripts and the output space of each branch, computed from the program.\\
$\mathcal{T}$ & \leanlink[https://maorbenshahar.github.io/QCryptLean/docs/QCryptLean/LOCC/Boundary.html\#LOCC.Boundary.Exit]{Exit} & The set of complete public histories, one for each branch; \leanlink[https://maorbenshahar.github.io/QCryptLean/docs/QCryptLean/LOCC/Transcript.html\#LOCC.Transcript]{Transcript} is the transcript register of a fixed list of alphabets.\\
denote & \leanlink[https://maorbenshahar.github.io/QCryptLean/docs/QCryptLean/LOCC/Program/Denotation.html\#LOCC.Program.denote]{denote} & Compiles a program into its channel, composing the branch Kraus maps and exposing announcements to Eve.\\
\addlinespace[3pt]
\colrule
\multicolumn{3}{l}{\textit{Security and classicality}}\\[3pt]
$\mathcal{E}^{\mathrm{real}}$ & \leanlink[https://maorbenshahar.github.io/QCryptLean/docs/QCryptLean/QKD/Protocol.html\#QKD.Protocol.real]{real} & The real channel $p.\mathsf{denote}$ determined by the protocol.\\
$\mathcal{E}^{\mathrm{ideal}}$ & \leanlink[https://maorbenshahar.github.io/QCryptLean/docs/QCryptLean/QKD/Protocol.html\#QKD.Protocol.ideal]{ideal} & $\mathcal{R}\circ\mathcal{E}^{\mathrm{real}}$, where $\mathcal{R}$ replaces both key registers with a single fresh uniform string (so the keys are equal, uniform, and independent of Eve) and leaves the transcript, other registers, and aborting runs unchanged.\\
key label & \leanlink[https://maorbenshahar.github.io/QCryptLean/docs/QCryptLean/QKD/KeyEnd.html\#QKD.KeyEnd]{KeyEnd} & Label carried by each $\mathsf{done}$ of a QKD program: abort, or accept $\ell$ with the part of Alice's and Bob's final registers that holds the $\ell$-bit key.\\
$\varepsilon$-security & \leanlink[https://maorbenshahar.github.io/QCryptLean/docs/QCryptLean/QKD/Protocol.html\#QKD.Protocol.IsSecure]{IsSecure} & Composable full-interface security: the accepted keys are classical and $\tfrac12\|\mathcal{E}^{\mathrm{real}}-\mathcal{E}^{\mathrm{ideal}}\|_\diamond\le\varepsilon$, comparing the real and ideal channels jointly over both keys, the transcript, and Eve's reference.\\
honest-classical & \leanlink[https://maorbenshahar.github.io/QCryptLean/docs/QCryptLean/LOCC/Program/TwoPartyClassicalStorage.html\#LOCC.Program.IsHonestClassical]{IsHonestClassical} & Every local action preserves diagonality on Alice's and Bob's registers, along every branch.\\
\end{tabular}
\end{ruledtabular}
\end{table*}

\bibliography{references}

\end{document}